# Characterization of magnetostatic surface spin waves in magnetic thin films: evaluation for microelectronic applications

Jae Hyun Kwon, Sankha Subhra Mukherjee, Praveen Deorani, Masamitsu Hayashi, and Hyunsoo Yang

*J. Kwon · S. S. Mukherjee · P. Deorani · H. Yang*
*Department of Electrical and Computer Engineering and NUSNNI-NanoCore, National University of Singapore, 117576, Singapore*

*M. Hayashi*
*National Institute for Materials Science, Tsukuba 305–0047, Japan*

The corresponding author:
H. Yang, eleyang@nus.edu.sg

S.S.M. and J.H.K. contributed equally to this work.

**Abstract** The authors have investigated the possibility of utilizing spin waves for inter- and intra-chip communications, and as logic elements using both simulations and experimental techniques. Through simulations it has been shown that the decay lengths of magnetostatic spin waves are affected most by the damping parameter, and least by the exchange stiffness constant. The damping and dispersion properties of spin waves limit the attenuation length to several tens of microns. Thus, we have ruled out the possibility of inter-chip communications via spin waves. Experimental techniques for the extraction of the dispersion relationship have also been demonstrated, along with experimental demonstrations of spin wave interference for amplitude modulation. The effectiveness of spin wave modulation through interference, along with the capability of determining the spin wave dispersion relationships electrically during manufacturing and testing phase of chip production may pave the way for using spin waves in analog computing wherein the circuitry required for performing similar functionality becomes prohibitive.

# 1. Introduction

Spin waves were first described by Bloch as collective, elementary excitations of individual magnetic moments, interacting with each other in a ferromagnet (FM). These excitations are precessional motion of spins, and propagate in the FM through exchange or magnetostatic interactions. Spin waves have been identified as promising candidates for information transfer [1, 2], quantum [3] and classical [4] information processing, control of THz dynamics [5] and phase-matching of nanotorque oscillators [6]. Ferromagnetic resonance and spin waves are used in the study of spin pumping [7-12]. They have also been used for the determination of damping using such techniques as spin torque induced ferromagneic resonance (ST-FMR) [13] as well as for the explanation of the spin Seebeck effect [14, 15].

In this work, we have investigated the possibility of using spin waves for data communication in microelectronic circuits, along with a study of its fesability in logic, using simulations as well as exeprimental measurements. Using simulations we have studied the effect of different material parameters on the spin wave propagation length as a means of determining the possibility of interchip communication using spin waves. For communications and logic, we have demonstrated methods for finding the dispersion relationship of spin waves in thin magnetic films via electrical measurements, so as to characterize the dispersion relationships of spin waves for applications in the manufacturing and testing stages. We have provided methods of determining wave vectors by applying a bias field partially out-of-plane of the sample in pulse inductive microwave magnetometry (PIMM) measurements [16]. Independent measurements of frequency and group velocity have been performed for NiFe films, and compared with theoretical calculations to verify the consistency of the adopted approach with theory. Wave properties of spin wave such as interference [17-20], reflection [21], or diffraction [22] have been reported in the previous studies. In our study, spin wave interference experiments have been performed in time domain using two pulses. We explore the possibility of using interference as a means of modulating transmitted signals for logic as well as a tool for amplification in communications.

## 2. Simulating the dynamics of spin waves

The characterization of the decay properties of spin waves, as they travel through the magnetic material, allows for an accurate determination of the damping of magnetization dynamics, an understanding of which is fundamental in many practical applications, such as the study of bit reversal in hard disk drives, the determination of the operational frequencies of spin torque oscillators [23], the determination of the field-time characteristics of switching in field-driven magnetic tunnel junctions (MTJ), and the determination of currents required for switching spin-torque-transfer MTJs [24]. For determining the damping characteristics, spin wave simulations have been performed using the OOMMF simulator [25]. This has allowed us to characterize not only how different materials behave, but also how changes in material parameters affect the attenuation length. In all our simulations, exchange, demagnetization, Zeeman, and magnetocrystalline energies are included in the simulations, while thermal fluctuations are neglected.

Snapshots of simulations performed on a 600 μm × 120 μm × 50 nm cuboidal sample with a cell size 50 nm × 6 μm × 50 nm are shown in Fig. 1(a). The $x$-axis lies along the length of the bar, and the $z$-axis points towards the reader in a right-handed Cartesian coordinate system. The parameters used in the simulation are as follows: Gilbert damping constant $\alpha = 0.01$, a saturation magnetization $M_s = 860 \times 10^3$ A/m, and an exchange stiffness $A = 1.3 \times 10^{-11}$ J/m. A bias field ($H_b$) of 100 Oe is applied along the $y$ direction, and thus the generated spin waves are magnetostatic surface waves (MSSW). Spin waves are generated in the middle of the strip at time $t = 0$ ns, and they progressively move away from the center with time. The intensity plot in Fig. 1(a) shows the $z$-component of the magnetization within the sample. In the simulations the Oersted field resulting from an antenna has been replicated using the Karlqvist equations [26] given by $H_x(x,z) = H_0 \left[ \arctan\left[(W/2+x)/z\right] + \arctan\left[(W/2+x)/z\right] \right]$, and $H_z(x,z) = \frac{H_0}{2} \left[ (W/2-x)^2 + z^2 \right] / \left[ (W/2+x)^2 + z^2 \right]$ where the time-dependent part of the Karlqvist field $H_0$ has a sinc-dependence to time, $I = I_0 \sin\left[\omega(t-t_0)\right] / \left[\omega(t-t_0)\right]$, with $\omega = 100$ GHz, $I_0 = 2 \times 10^{-4}$ A, and $t_0 = 50$ ns. The waveguide width $W$ is 2 μm. $H_0$ and $I$ are related as $H_0 = I/2W$.

Thus the amplitude of the magnetic field pulse is about 7 Oe in the *x* direction and 6.1 Oe in the *z* direction, as shown in Fig. 1(b).

Spin wave propagation loss is one of the critical factors required for modeling of spin wave devices. These losses determine the distance which the spin waves can travel before they become too small to be detected. There have been both experimental and theoretical studies of propagation loss [27-32]. A phenomenological loss theory determines the upper and lower limits of propagation losses for different propagation modes of MSSW [32, 33]. However, an accurate description of magnetization dynamics is very difficult as the boundary conditions and initial conditions are difficult to access. In order to avoid this problem, numerical methods such as finite-differential method (FDM) or finite-element method become very helpful. In this regard, a study of spin wave propagation losses based on numerical simulations is of direct importance.

In this study, micromagnetic simulations are used to study the attenuation characteristics of MSSW. Simulations are done with cell a size 50 nm × 120 μm × 50 nm on a 600 μm × 120 μm × 50 nm cuboidal sample. Spin waves are excited from a waveguide located at the center of the sample. The waveguide has a width of 2 μm and a thickness of 200 nm, and is separated from the sample by a 50 nm thick insulator. A bias field ($H_b$) of 100 Oe is applied along the *y* direction. A sinc pulse with a frequency of 100 GHz is used to generate a pulse field in the sample with the spatial profile given by Karlqvist equations. The reason for using a sinc pulse at 100 GHz is that in the frequency domain, this pulse has a uniform distribution in 0-15 GHz. The amplitude of the magnetic field pulse is the same as that in Fig. 1(b).

The spin wave amplitude, defined as the maximum variation in the *z*-component of magnetization at different simulation times, is measured at different locations in the sample and plotted against the distance from the source of the spin waves. The data, between 7 and 77 μm from the source are fitted with an exponentially decaying function to obtain the attenuation length of the spin waves as shown in Fig. 2. The attenuation length ($l_{Att}$) is defined as the distance the wave travels during which its amplitude decreases by 1/e. The parameters used for the simulations of the different materials are shown in Table 1. For $CoFe_2Al$ and GaMnAs, uniaxial anisotropy along the *z* direction has been included in the simulations. The attenuation lengths are 11.54, 26.32, 18.95, and 3.42 μm for permalloy (Py), yttrium-iron-garnet (YIG), CoFe$_2$Al, and GaMnAs, respectively. It is noteworthy that the spin waves in all these materials propagate through similar distances. However, the amplitude of spin wave is found to be of similar order in Py, YIG, and CoFe$_2$Al, while in GaMnAs it is about four orders of magnitude smaller. This result explains the difficulties associated with observation of propagating spin waves in dilute magnetic semiconductors (DMS). The challenge of measuring such small amplitude spin waves is a constraint that needs to be considered when designing a spin wave device based on DMS. The small amplitude of spin waves in GaMnAs is due to the small value of saturation magnetization ($M_s$ = 40×10$^3$ A/m) and a high value of damping constant ($\alpha$ = 0.028).

In these simulations, the cell size in the $y$ direction is 120 μm, which is much larger than the spin wave decay lengths. Since the waves travel only in the $x$ direction, and the *sinc* pulse has components in the $x$ and $z$ directions, we do not expect any dynamics in the $y$ direction and hence the cell size in the $y$ direction should not have any effect on magnetization dynamics. In order to confirm this, simulations were performed with cell size 50 nm × 500 nm × 50 nm with parameters for Py ($\alpha$ = 0.01, $M_s$ = 860×10$^3$ A/m, and A = 1.3×10$^{-11}$ J/m), and identical results to the larger cell size (50 nm × 120 μm × 50 nm) were obtained.

We further extend the simulations by varying different parameters such as the damping constant ($\alpha$), saturation magnetization ($M_s$), bias field ($H_b$), and exchange stiffness ($A$) to see the effect of these parameters on spin wave attenuation length. For each set of simulations only one parameter is changed and the other parameters are kept the same as those of Py. Figure 3(a) illustrates the effect of damping constant ($\alpha$) on the attenuation length ($l_{Att}$). As $\alpha$ decreases, initially the attenuation length increases logarithmically and then becomes nearly constant. This behavior is explained by considering the two main factors responsible for attenuation of spin wave amplitude: (1) energy loss by various damping mechanisms represented by $\alpha$, and (2) spreading of the wave packet as it travels down the film due to a nonlinear dispersion relationship. For higher values of $\alpha$, the energy loss mechanisms are dominant, whereas for lower values of $\alpha$, dispersion becomes the dominant factor in determining the spin wave amplitude. Figure 3(b) shows the dependence of $l_{Att}$ on the saturation magnetization. In the range $M_s$ = 60×10$^3$ – 1800×10$^3$ A/m, $l_{Att}$ increases nearly logarithmically with increasing $M_s$ and saturates at higher values of $M_s$. Figure 3(c) shows that $l_{Att}$ decreases as $H_b$ increases when $H_b > 20$ Oe. However, for $H_b < 20$ Oe, $l_{Att}$ increases with increasing $H_b$. This is probably because spin waves are non-linear for $H_b < 20$ Oe. In Fig. 3(d) we show that $l_{Att}$ does not change with the exchange stiffness parameter. This is expected since, in our simulations, the waveguide is very wide (2 μm), therefore, the waves are magnetostatic spin waves rather than exchange-coupled waves.

It is remarkable that $l_{Att}$ does not change very much, even when the parameters ($\alpha$, $M_s$, $H_b$ and $A$) are varied over wide ranges. It means that choosing the right material is not an efficient method for tuning the propagation length of spin waves. However, as mentioned earlier, it is ultimately the amplitude of the spin waves that is measured in real experiments, and in this regard the choice of materials becomes very important. As described earlier, there are two reasons for attenuation of spin waves – damping and dispersion. The second factor can be eliminated, if a sinusoidal field is used instead of a pulse field to excite spin waves. In order to confirm this, simulations have been done with a sinusoidal field at the resonance frequency of ferromagnets. All the other details of the simulations remain same. The $l_{Att}$ is found to be 31.92 μm and 1910.4 μm for Py and YIG, respectively. These values are significantly larger for sinusoidal fields than the sinc fields, and explain the observation of spin waves a few tens of μm away from point of excitation in Py, and even hundreds of μm in YIG [2, 34, 35].

As can be seen from the simulations, the attenuation characteristics of YIG allow for the transmission of spin waves over several millimeters with very little loss, as shown previously [2, 34]. However, the fabrication of YIG films is not compatible with concurrent complimentary metal oxide semiconductor (CMOS) technology [36]. Thin-films such as Py which can be easily deposited are preferable, however, the propagation length is much shorter than what is required for chip-to-chip communications. Hence, it is practically impossible to use spin waves for interchip communications, but intrachip communications and spin wave-based logic may still be possible. For this reason, we have experimentally evaluated all-electrical ferromagnetic thin-film magnonic systems to determine their applicability for intrachip communication in the following section.

## 3. Wave properties of spin waves

Spin waves exhibit a number of different wave properties, the most fundamental of which is the transfer of information from one point to another. Electrical spin wave sources and detectors comprise of microwave antenna which can generate high-frequency Oersted fields or those which can inductively detect the same. For practical applications all-electrical systems are required for intra-chip communications and logic. In this section, we have examined the wave properties of spin waves in thin ferromagnetic films.

As is well known, dispersion characteristics of waves are fundamental to the determination of the transmission characteristics of waves. Although the determination of precessional frequency using purely electrical means is relatively easy, the determination of *k*-vectors is rather challenging. In section 3.1, electrical sources and detectors have been used for extracting the dispersion relationship in spin waves. Furthermore, spin waves used for communications and logic in electronic circuits would necessarily benefit from pulsed electrical measurements because a vast majority of modern electrical circuits are digital circuits. Pulse inductive microwave magnetometry (PIMM) is thus a very natural measurement procedure for such intrachip communications. However, since the attenuation is significant, a method for the modulation of spin waves in PIMM measurements is very beneficial. In section 3.2, a method for achieving such modulation in PIMM measurements has been demonstrated, which may be useful in combating such problems with attenuation.

### 3. 1. Spin wave dispersion

The dispersion relationship of spin waves is strongly correlated to the applied field, because the field changes the characteristics of the material through which the spin waves travel. The spin waves are characterized by the precessional frequency of the magnetization and the wavelength. In the current measurements, we have used PIMM [37, 38] for studying the magnetization dynamics. In particular, the wave-vectors of travelling spin wave packets resulting from impulse excitations are studied, and their wave-vectors are extracted. Since PIMM is a relatively fast and useful method of measuring magnetization dynamics, it would be convenient to have a relatively easy method for extracting the wave-vectors from these measurements electrically.

As shown in Fig. 4(a) and (b), two devices have been used for the extraction of the wave vectors of travelling spin waves. For the device in Fig. 4(a) a 120 μm × 200 μm Ni$_{81}$Fe$_{19}$ (30 nm) pattern is covered by a 30 nm thick SiO$_2$ insulating layer, and finally Cr (5 nm)/Au (100 nm) ground-signal-ground (GSG) waveguides, with 8 μm signal lines are patterned on top. The device in Fig. 4(b) comprises of a 220 μm × 340 μm Ni$_{81}$Fe$_{19}$ (20 nm) pattern, covered by a 30 nm thick SiO$_2$ layer, with asymmetric coplanar strips (ACPS) patterned on top with 10 μm signal lines. A 2 V, 100 ps pulse is applied and the Gaussian wave packet detected at the detector waveguide, in the device shown in Fig. 4(a) is shown in Fig. 4(c) with a 215 Oe bias field applied parallel to the film plane, along the signal line. The angular frequency $\omega$ of the measured signal, and the applied bias field ($H_b$) are related by the dispersion relation of MSSW: $\omega^2 = \gamma^2 \mu_0^2 \left[ H_b(H_b + M_S) + M_S^2 (1 - e^{-2kd})/4 \right]$ . Here, $\gamma$ represents the gyromagnetic ratio of an electron in free space, $\mu_0$ the permeability of free space, $M_s$ the saturation magnetization ($\mu_0 M_s \approx$ 1 T for NiFe), $d$ the sample thickness, $k$ the wave number (=2π/λ), and $\lambda$ is the wavelength of the spin wave. However, it is generally relatively difficult to extract the value of $\lambda$ from electrical measurements because these measurements are position-independent. When the magnetic field is applied out of the plane of the sample, and is less than the field required to saturate film, the in-plane component of the magnetic field is still used for producing surface waves.

In Fig. 5, the frequency transform of the measured time-domain signals of the sample shown in Fig. 4(a) are shown. For bias fields applied partially out-of-plane and taking the FFT of these signals, the resultant frequency of the measured spin waves is seen to shift steadily from a higher to a lower frequency with increasing Θ, under similar magnetic bias conditions as shown in Fig. 5(a-e). The width of the signal line has been commonly used for calculating the wave number $k$. As previously shown by optical measurements, generated spin waves have wavelengths between the width of the signal line and infinity (i.e. $k$ = 0) [39, 40]. As can be seen, the frequency response falls within the dispersion regimes given by the surface wave equation $\omega\left(k = \dfrac{2\pi}{8 \mu m}\right)$ shown as thick dashed lines, and the ferromagnetic resonance (FMR) mode $\omega(k=0)$ shown as thin dashed lines. At $k = 0$, the waves do not travel, and represent FMR spectra. For FMR, the resonant frequency is zero at no applied bias. Note that in fitting equations $H_b$ has been replaced with the in-plane component of the field $H_b \cos(\Theta)$. Note that when the applied magnetic field is in the low-bias field regime in which the measurements have been performed, even though the field is out of the plane of the film, the measured spin waves still faithfully reproduce the dispersion curves of surface spin wave mode. A yellow line shows the fits to the magnetostatic backward volume (MSBV) mode $\omega^2 = \gamma^2 \mu_0^2 \left[ H_b \left( H_b + M_S \left( 1 - e^{-kd} \right) / kd \right) \right]$. As can be seen, the calculated frequencies are lower than that required for explaining the observed behavior. Furthermore, the equation for the magnetostatic forward volume (MSFV) mode is only applicable, when the applied bias field is greater than the saturation magnetization. Hence, for the range of fields which have been applied, the MSFW mode is not generated.

In order to support this experimental data, simulations were performed with $H_b$ applied at an angle $\Theta$ out of the plane of the sample. The material parameters used in the simulation are the same as used for permalloy earlier, and $H_b$ is taken to be 1 kOe. The geometry of sample used for these simulations is similar as that used earlier for study of $l_{Att}$. In these simulations, we have studied the variation of spin wave frequency as a function of $\Theta$. The normalized frequency with respect to that at $\Theta = 0$ is shown as red squares in Fig. 5(f). It is then fitted with frequencies calculated from the MSSW dispersion relation with $k = 2\pi/(10 \ \mu m)$, in which 10 μm is the antenna width of the signal line. These calculated frequencies, after normalizing with respect to the maximum, are shown as black line in the figure. Using theoretical fits of the surface wave dispersion relationship to both measured (Fig. 5(a-e)) and simulated data (Fig. 5(f)), one can conclude that the spin wave mode propagating, when the field is applied partially out of the plane of the sample, is essentially the surface wave.

Using this information, a detailed mapping of the prevalent spin wave modes may be extracted. For this purpose, the sample shown in Fig. 4(b) has been used, wherein the applied magnetic field has been applied out of the sample plane. Smaller waveguides are able to measure the signal levels to much greater accuracy, and individual lines corresponding to different $k$ vectors may be extracted, as shown in Fig. 6(a). From the fitting dispersion curves corresponding to the wavelengths 10, 14, 23, and 50 μm have been represented with a dotted line, dash-dotted line, dashed line, and a thin solid line, respectively. The measured frequency responses also have wavelengths between 10 μm and the FMR spectrum (thick solid line) respectively.

As an additional check to ensure that the calculations are consistent, the theoretical group velocity ($\partial_k \omega$) for the different measured $k$ vectors is compared with measured values of group velocity, using two devices having different excitation-detection distances of 20 μm (device $D_{20}$) and 5 μm (devices $D_5$). The wave packets are fitted with the Gaussian function $A \exp\left((t-t_0)^2 / 2\sigma^2\right)$, with fitting parameters $t_0$, $\sigma$, and $A$, for bias fields between 25 and 120 mT [41]. Only the temporal position ($t_0$) of the center of the wave packet has been obtained by fitting a Gaussian function to the wave packet. The measured group velocity is defined as $(20\,\mu m - 5\,\mu m) / \left(t_0|_{D_{20}} - t_0|_{D_5}\right)$. Notice that this is a direct measure of the speed of the Gaussian spin wave packets, and is independent from the measurements of the frequency of the individual packets. These are shown in Fig. 6(b) and show very good correspondence with predicted theory. The theoretical group velocity for the different wave vectors shown by cyan lines in Fig. 6(a) has been calculated by the formula

$$v_g = \frac{d\omega}{dk} = \frac{\gamma^2 \mu_0^2 M_S^2}{4}\left(\frac{d}{\omega}\right) e^{-2kd} .$$

As can be seen from Fig. 6(a), at low values of the bias magnetic field, the spin wave a wavelength of 50 μm is dominant. As the bias magnetic field increases the spin waves with a wavelength of 50 μm slowly dwindle, and those with a wavelength of 23 μm become more pronounced. As the field increases further the signal to noise ratio slowly degrades and it becomes increasingly difficult to fit Gaussian wave functions for the data. As can be seen in Fig. 6(b), at low values of the applied bias magnetic field, the velocity corresponds to spin waves having a wavelength of 50 μm. As the bias magnetic field increases, however, the measured group velocity also shifts toward that resulting from spin waves with a wavelength of 23 μm. Thus, one observes close agreement between values of group velocity measured experimentally and theory, confirming again that the observed spin waves are indeed surface waves. Note that two measurements in Fig. 6(a) and 6(b) are two independent techniques, and both are in good agreement with one another. Hence, one can conclude, that the method described in this section is a practical and useful method of extracting the *k*-vectors of spin waves in magnetic materials embedded in electronic circuitry.

### 3.2. Spin wave interference

The microscope image of the sample used for the spin wave modulation studies is shown in Fig. 7(a). A Py thin film was deposited and patterned on an MgO substrate with dimensions of 410 × 220 μm$^2$ and a thickness of 20 nm. Asymmetric coplanar waveguides (CPW) are sputter deposited on the patterned Py and is isolated from Py by 30 nm of SiO$_2$. The signal line width of the CPW is 10 μm. The gap between the signal line and the ground conductor is 5 μm, and the distance between two CPWs is 30 μm. PIMM is used to generate and detect spin wave packets [16, 42]. A pulse generator is used to apply an impulse excitation at the input CPW and the resultant Gaussian wave packet is detected via a low noise amplifier by a 50 GHz sampling oscilloscope. Travelling spin wave packets are detected in the MSSW mode, in which spin wave *k*-vector is perpendicular to the direction of magnetization, **M** of the ferromagnetic thin film. The contour plot in Fig. 7(b) shows color coded amplitudes of spin wave signals excited by a single pulse at different bias fields. The dashed line in the inset of Fig. 7(b) indicates spin wave packet at the fixed $H_b$ of 41 Oe. The bias field dependent shift of the resonance frequency, $f_R$, obtained by taking a fast Fourier transform (FFT) of the time-domain signals is shown in Fig. 7(c). The quadratic dependence of $f_R$ with increasing $H_b$ is shown in Fig. 7(c) by a dashed line [43].

Two spin waves excited by two pulses can interfere either constructively or destructively, where one pulse is fixed and the other is shifted by a time delay (Δ*t*) indicated by the arrows in Fig. 7(d). The time delay between the two pulses is changed from +5 ns to -5 ns in steps of 20 ps, while the bias field $H_b$ is fixed at 123 Oe. Two different measurements were performed, and are termed bipolar and unipolar experiments, respectively. In the bipolar experiment, the two impulses have opposite polarity, whereas, in the unipolar experiment, both pulses have the same polarity. Contour plots of the spin wave interference by the bipolar and unipolar pulses with different time delays are shown in Figs. 8(a) and 8(c), respectively. The blue dashed lines indicate constructive interferences at Δ*t* = ±(2n+1)139 ps for the bipolar pulse, and ±(2n)139 ps for the unipolar pulse, where n is an integer number. The red dashed lines indicate destructive interferences at $\Delta t = \pm(2n)139$ ps and $\pm(2n+1)139$ ps for the bipolar and unipolar, respectively, as shown in Figs. 8(b) and 8(d). When two spin waves constructively interfere, the amplitude of interfered spin waves is the sum of the amplitudes of each spin wave generated by the independent impulses. The amplitude of the non-interacting wave packet for bipolar (unipolar) experiment is ~1.7 mV (~1.63 mV). When they constructively interfere, the amplitude is ~3.4 mV (~3.3 mV) in case of bipolar (unipolar) input pulses. Signals interfering destructively have a magnitude of almost zero. Constructive and destructive interference caused by a linear superposition of two spin waves [20] can be utilized for possible applications in spin wave logic devices.

FFT is used to obtain further insight of the phenomenon in the frequency domain. Color-coded frequency domain spectra are shown in Figs. 9(a) and 9(c), as a function of $\Delta t$, for bipolar and unipolar pulses, respectively. As shown in Fig. 7(c), the resonance frequency, $f_R$ is ~3.59 GHz at a bias field $H_b$=123 Oe. The dashed squares in the middle of Fig. 9(a) and 9(c) are expanded in the respective insets. The normalized FFT amplitude in the frequency domain changes periodically with a periodicity of ~278 ps. The horizontal dashed lines in Fig. 9(a) and 9(c) are the $\Delta t$-dependent changes in FFT amplitude, extracted at $f_R$=3.59 GHz and are plotted in Fig. 9(b) and 9(d), respectively. The spin wave packets that interfere due to unipolar and bipolar pulses can be described in time domain as

$$V_{uni}(t) = A\cos(2\pi f_R t)\exp(-\frac{t^2}{2\sigma^2}) + A\cos[2\pi f_R(t-\Delta t)]\exp(-\frac{(t-\Delta t)^2}{2\sigma^2}) \text{, and}$$

$$V_{bi}(t) = A\cos(2\pi f_R t)\exp(-\frac{t^2}{2\sigma^2}) + A\cos[2\pi f_R(t-\Delta t)+\pi]\exp(-\frac{(t-\Delta t)^2}{2\sigma^2}),$$

respectively, where $\sigma$ is the full width at half maximum of a single wave packet in time domain. The interfered amplitude of both the unipolar and bipolar experiments, $V_{uni}(t)$ and $V_{bi}(t)$, is the summation of the wave packet amplitudes excited by the fixed pulse and the shifting pulse, which are the first and second terms in the equations shown above, respectively. The wave packet excited by the bipolar impulse is phase shifted by $\pi$ with respect to that excited by unipolar pulses [19]. After applying FFT, as can be seen in the Fig. 9(b) and 9(d), the intensity of the interference varies sinusoidally with $\Delta t$. The normalized FFT amplitude changes periodically with a periodicity of 278 ps. Note that 278 ps is equal to $1/f_R$ (=1/3.59 GHz). Complete constructive or destructive interference indicates 1 and 0, respectively, in the normalized FFT amplitude. In the unipolar experiment, two wave packets interfere either constructively at $\Delta t$=0 or destructively at $\Delta t$=±139 ps. In the bipolar case, the destructive and constructive interferences are observed at $\Delta t$=0 and $\Delta t$ =±139 ps, respectively. The amplitudes for $|\Delta t|$ > 5 ns, where the interference becomes negligible, are close to half the value of the maximum FFT amplitude. Hence, we can use this method to modulate the intensity of the spin waves through interference. This can be effectively used for engineering spin wave intensity for communication and logic.

## 4. Conclusion

We have studied how different parameters affect the spin wave attenuation length through micromagnetic simulations. We have shown that the attenuation is most strongly affected by the damping constant above a certain value, below which other means of attenuation such as that resulting from nonlinear dispersion become more dominant. Since the attenuation length of spin waves are of the order of several microns for most manufacture-friendly material systems, the possibility of inter-chip communication using spin waves is ruled out. A few tens of microns is generally sufficient for intra-chip communications. We have also demonstrated a technique for determining the dispersion relationships by electrical methods which would aid in the determination of the propagation characteristics of spin waves in applications. Finally, we have also demonstrated spin wave interference using electrical techniques.

Spin wave computation has been explored by two different techniques. First, Khitun *et al*. [44] have attempted to implement logic using the spin wave bus. However, they have used the phase of spin waves for implementing logic circuits, which makes it difficult for use in logic circuits, because the phase is a continuous variable and not a discrete binary one. The approach taken by Kolokoltsev *et al.* [45] is much more practical, in which a PSK signal has been synthesized using spin waves. The digital circuitry required for implementing PSK is significant, and hence, such analog implementations may be more appropriate for spin wave logic. Using the experimental implementations described above, it may thus be possible to augment digital logic with efficient spin wave-based analog computation.

Acknowledgements

This work is supported by the Singapore National Research Foundation under CRP Award No. NRF-CRP 4-2008-06 and Grant-in-Aid for Scientific Research (No. 22760015) from MEXT, Japan.

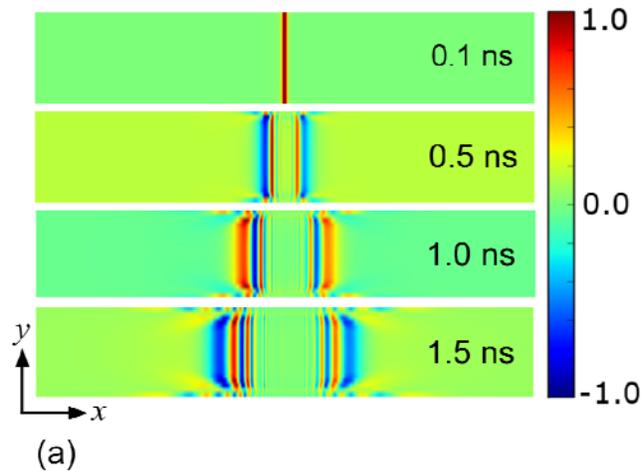

(a)

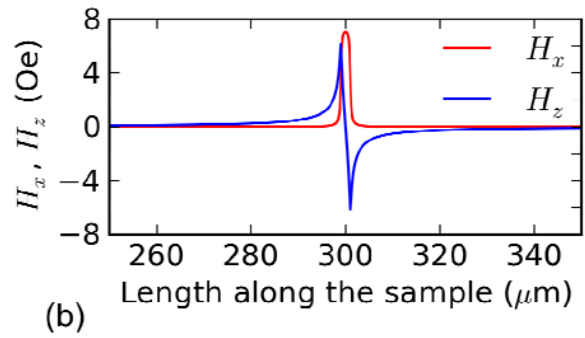

(b)

**Fig. 1.** (a) The normalized *z*-component of magnetization in the sample plotted for different simulation times is shown, depicting the the generation and propagation of surface waves with time. (b) The magnetic field distribution used for simulating the Oersted field.

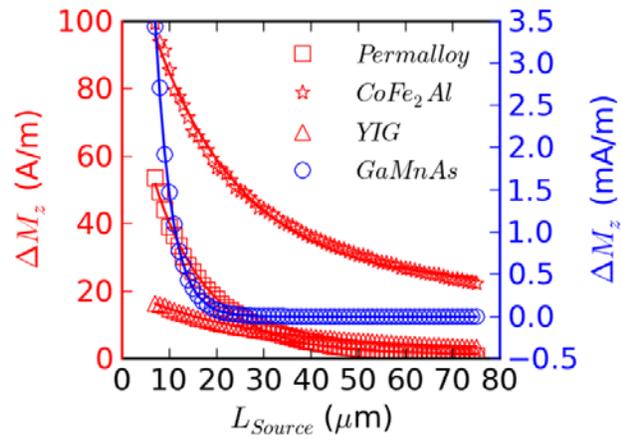

**Fig. 2.** The decay in the *z*-component of the magnetization as a function of the distance from the source has been plotted for different materials using open symbols. Solid lines represent exponential fits to the experimental data, used for extracting the decay length of the travelling waves. In GaMnAs the amplitude of spin waves is 4 orders of magnitude smaller, so the data for GaMnAs corresponds to the right axis with smaller units (mA/m).

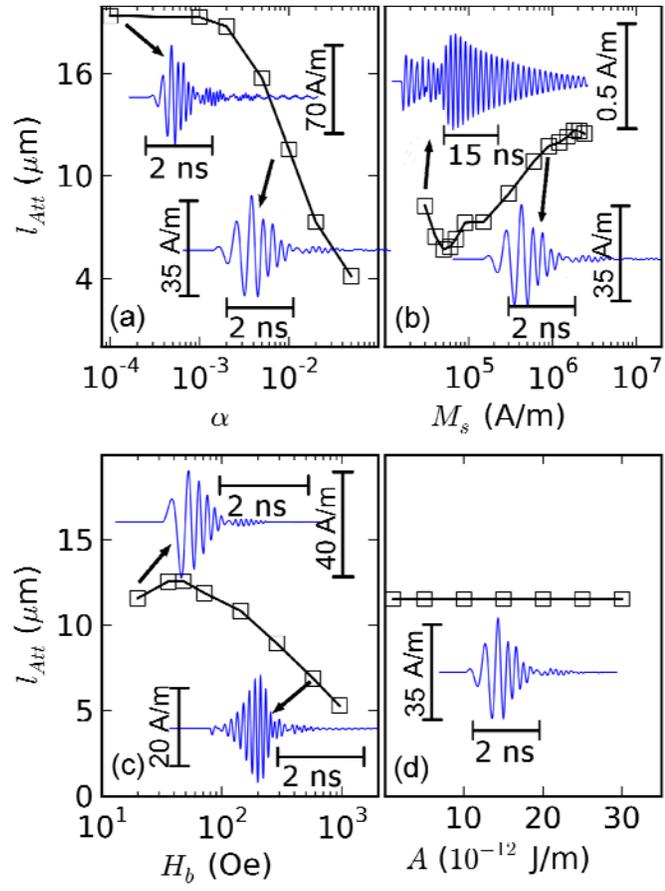

**Fig. 3.** The attenuation length ($l_{Att}$) of spin wave packets is shown as a function of the damping constant ($\alpha$) in (a), the saturation magnetization ($M_s$) in (b), the bias magnetic field ($H_b$) in (c), and the exchange stifness constant ($A$) in (d).

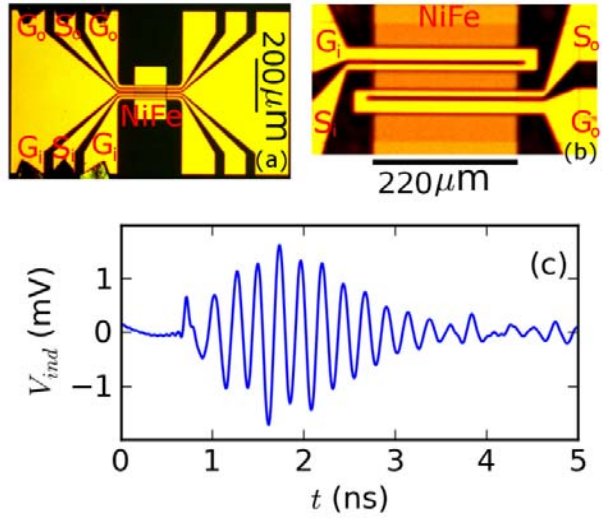

**Fig. 4** The two devices used for determining the *k*-vectors are shown. (a) Input and output Cr/Au GSG waveguides are deposited over a rectangular NiFe pattern. One of the grounds of the input and output probes are connected. (b) Ta/Au ACPS strips are patterned over a rectangular NiFe pattern. In both (a) and (b), subscripts 'i' and 'o' refer to input (excitation) and output (detection) waveguides, respectively, and the waveguides are separated from the NiFe strip by an insulating oxide layer. (c) The inductive spin wave signal obtained for an in-plane bias field of 215 Oe for the device shown in (a).

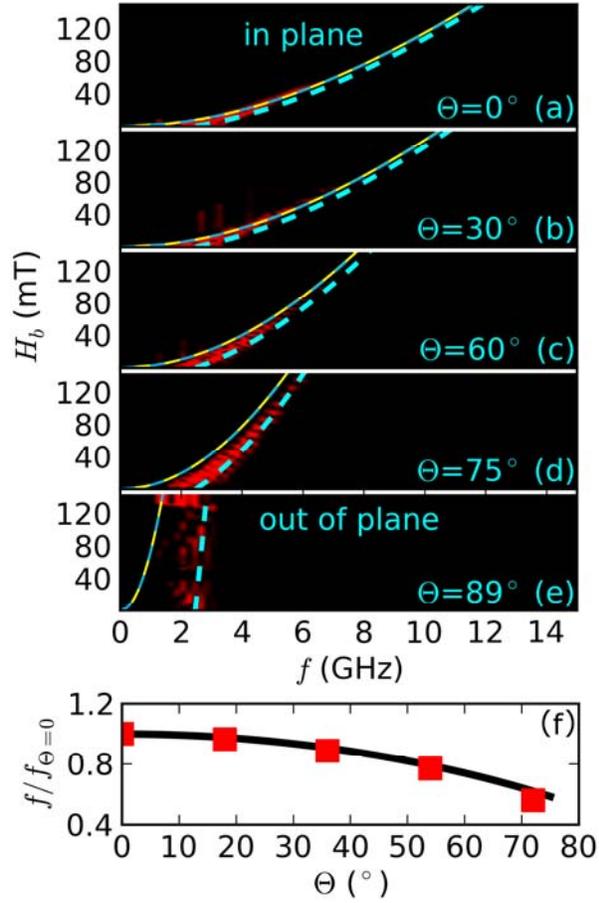

**Fig. 5**. (a-e) The FFT of time-domain Gaussian wave packets obtained from PIMM measurements on the device shown in Fig. 4(a), as a function of the applied bias field. The bias field is applied at an angle Θ out of the sample plane. The in-plane component of the bias field always points along the hard axis of the NiFe pattern (i.e. pointing from $G_i$ to $S_i$ in the figure). (f) The frequencies (normalized to that at Θ = 0) obtained from simulations for field applied at angle Θ out of the sample plane is (red squares) is compared with the equation
( $\cos(\Theta)(\cos(\Theta)+10.807)+0.25\times10.807^2(1-\exp(-8\pi\times10^{-3}))$ )$^{0.5}$ after normalization, wherein the simulations are performed at $H_b$ = 1 kOe and $M_s$ is taken to be 10.807 kOe, showing excellent match between simulation and theory.

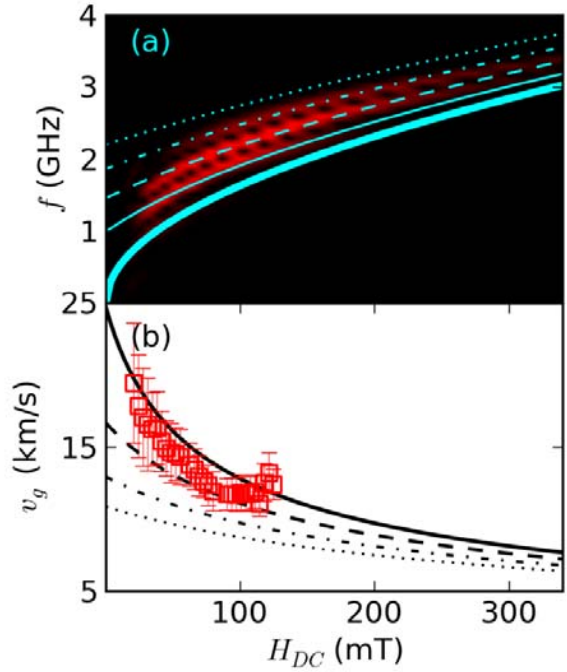

**Fig. 6**. (a) The FFT of time-domain Gaussian wave packets obtained from PIMM measurements on the device shown in Fig. 4(b) is shown as a function of the applied bias field. Calculations of spin wave frequencies for different wavelengths are shown as a function of applied bias. The thick solid line represents the FMR spectrum. (b) The calculated values of group velocity are plotted for spin waves with different wavelengths, as a function of the bias field. The measured values (symbols) of group velocity are overlaid on these plots, and show strong correlation with calculations. The wavelengths 10, 14, 23, and 50 μm have been represented with a dotted line, dash-dotted line, dashed line, and a thin solid line, respectively.

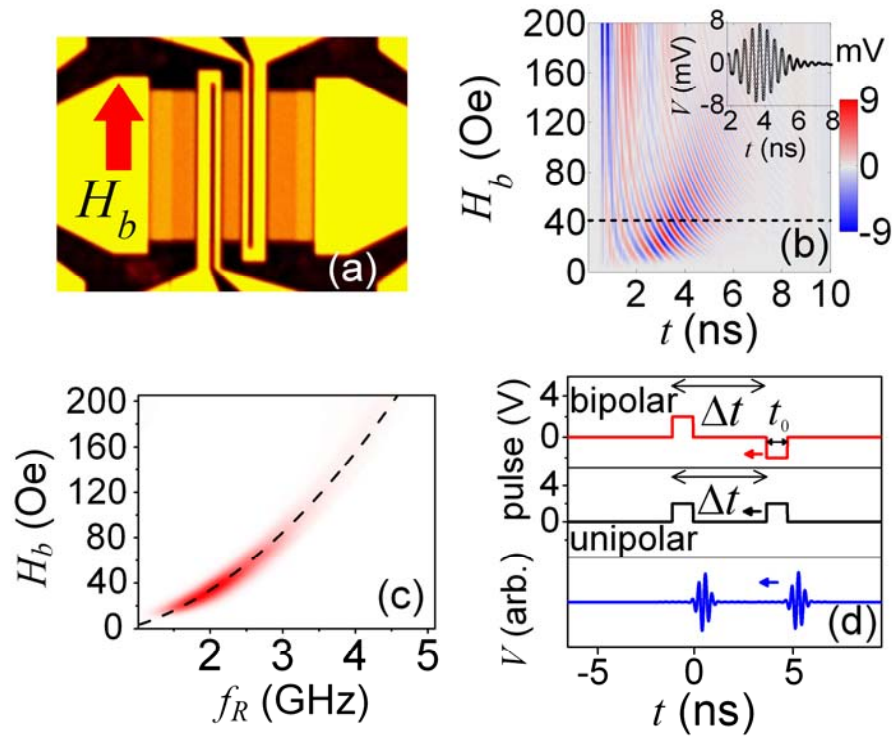

**Fig. 7.** (a) Two asymmetric coplanar waveguides are used for excitation and detection of spin wave packets. The wave guides are on top of FM and separated from FM by the insulator. The bias field, $H_b$, is along the waveguide direction. (b) Contour plot of spin waves induced by a single pulse is shown in time domain as a function of $H_b$. The signals become weak as $H_b$ increase. The spin wave packet at $H_b$ of 41 Oe is shown in the inset. (c) FFT of time domain signal shows the resonance frequency ($f_R$) as a function of $H_b$, fitted by the dashed line using the Kittle formula. (d) Two impulses separated from one another by $\Delta t$ is shown to produce two Gaussian wave packets.

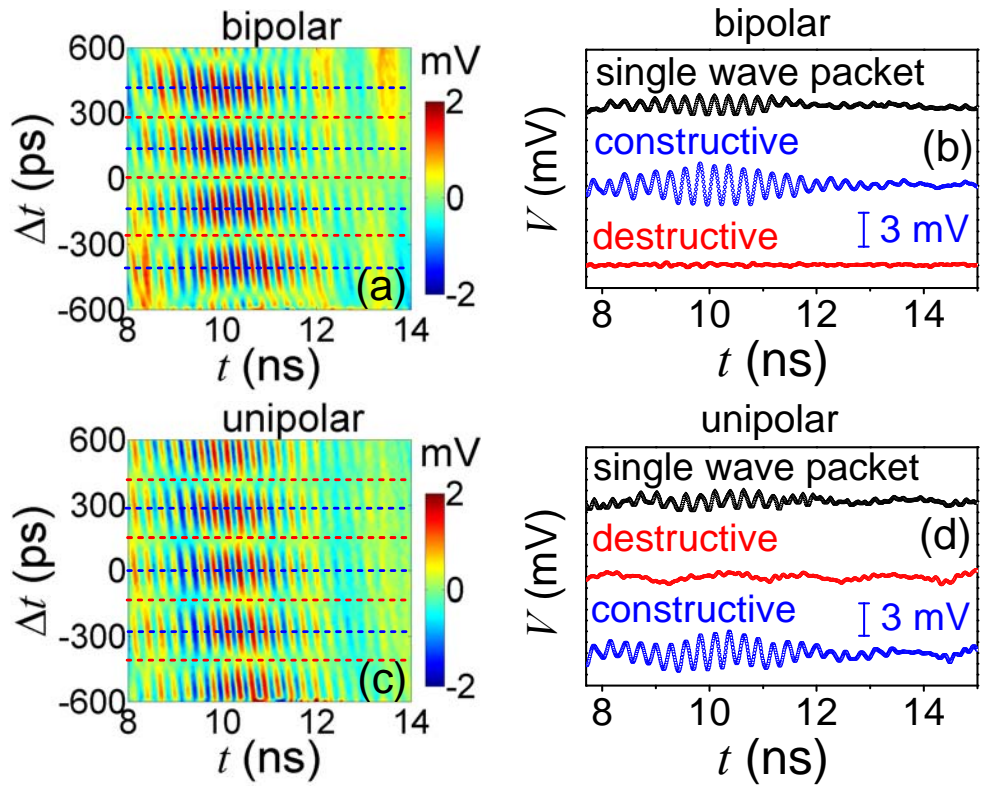

**Fig. 8.** (a), (c) Contour plots of spin wave interference for the bipolar and unipolar measurement, respectively, in time domain. Constructive (dashed blue line) and destructive (dashed red line) interference of (a), (c) are shown in (b), (d) for the bipolar and unipolar cases, respectively, where $\Delta t$ is either 0 or +139 ps.

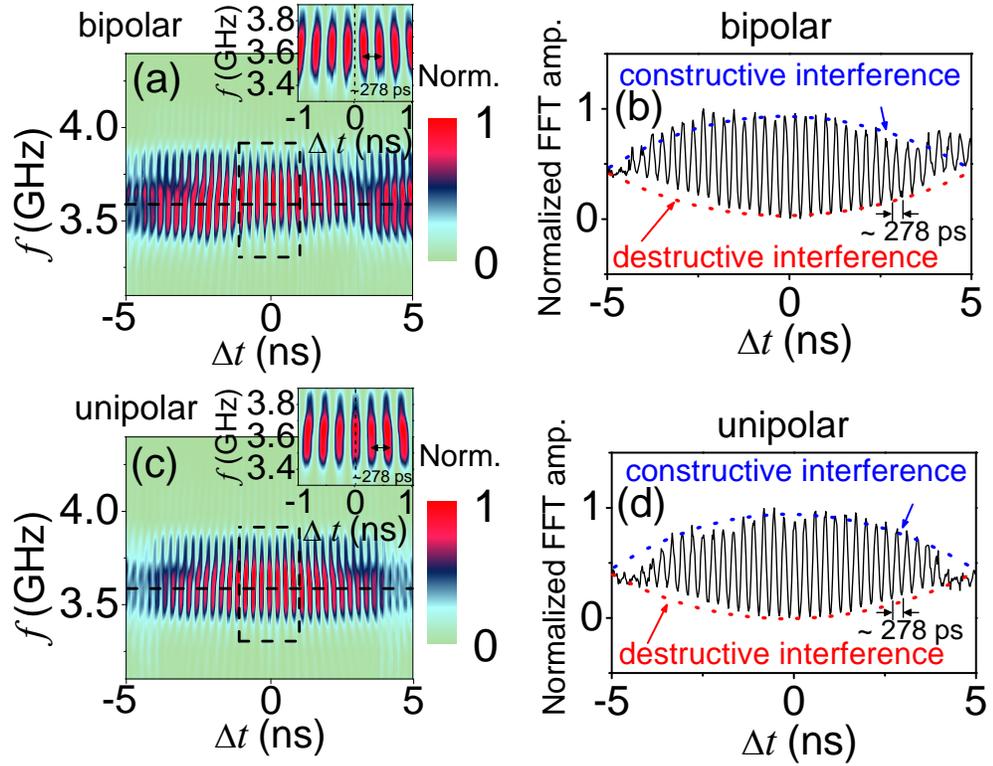

**Fig. 9.** (a), (c) show the contour plots of the frequency spectrum of spin wave in the bipolar and unipolar experiment, respectively. (b), (d) The normalized FFT amplitude at $f_R$=3.59 GHz is shown. The FFT amplitude without any interference ($\Delta t \geq \pm 5$ ns) is half of the amplitude when interfering constructively at $\Delta t$ =0 ps (unipolar) or $\pm$139 ps (bipolar). The FFT amplitude is 1 or 0, when two spin waves interfere constructively or destructively, respectively.

**Table 1** Parameters used for the different materials in the simulations.

| Material | Damping constant | Saturation magnetization (A/m) | Exchange stiffness (J/m) | Magnetocrystalline anisotropy (J/m$^3$) | Ref. |
|---|---|---|---|---|---|
| Py | 0.01 | 860×10$^3$ | 1.3×10$^{-11}$ | - | |
| YIG | 0.000067 | 150×10$^3$ | 4.2×10$^{-12}$ | - | |
| CoFe$_2$Al | 0.001 | 1053×10$^3$ | 1.5×10$^{-11}$ | Uniaxial, -1000 | [46] |
| GaMnAs | 0.028 | 40×10$^3$ | 2.24×10$^{-13}$ | Uniaxial, - 4000 | [47-49] |